# Friction decoupling and loss of rotational invariance in flooding models

Luca Cozzolino[1], Giada Varra[2], Luigi Cimorelli[3], Domenico Pianese[4], Renata Della Morte[5]

**Abstract**

*Friction decoupling, i.e. the computation of friction vector components making separate use of the corresponding velocity components, is common in staggered grid models of the SWE simplifications (Zero-Inertia and Local Inertia Approximation), due to the programming simplicity and to the consequent calculations speed-up. In the present paper, the effect of friction decoupling has been studied from the theoretical and numerical point of view. First, it has been found that friction vector decoupling causes the reduction of the computed friction force and the rotation of the friction force vector. Second, it has been demonstrated that decoupled-friction models lack of rotational invariance, i.e. model results depend on the alignment of the reference framework. These theoretical results have been confirmed by means of numerical experiments. On this basis, it is evident that the decoupling of the friction vector causes a major loss of credibility of the corresponding mathematical and numerical models. Despite the modest speed-up of decoupled-friction computations, classic coupled-friction models should be preferred in every case.*

[1] Ass. Prof., Ph. D., Dept. of Engrg., Parthenope Univ., Centro Direzionale di Napoli – Is. C4, 80143 Napoli, Italy. E-mail: luca.cozzolino@uniparthenope.it
[2] Ph. D. Stud., Dept. of Engrg., Parthenope Univ., Centro Direzionale di Napoli – Is. C4, 80143 Napoli, Italy. E-mail: giada.varra@uniparthenope.it
[3] Sen. Res., Ph. D., Dept. of Civil, Architectural and Environmental Engrg., Federico II Univ., via Claudio 21, 80125 Napoli, Italy. E-mail: luigi.cimorelli@unina.it
[4] Full Prof., Dept. of Civil, Architectural and Environmental Engrg., Federico II Univ., via Claudio 21, 80125 Napoli, Italy. E-mail: pianese@unina.it
[5] Full Prof., Ph. D., Dept. of Engrg., Parthenope Univ., Centro Direzionale di Napoli – Is. C4, 80143 Napoli, Italy. E-mail: renata.dellamorte@uniparthenope.it

**Subject headings*:* Friction; Flooding; Shallow-water Equations; Zero-Inertia model; Local-Inertia model; Rotational invariance.

Corresponding author: Luca Cozzolino

E-mail address: luca.cozzolino@uniparthenope.it

Address: Dipartimento di Ingegneria, Università degli Studi di Napoli Parthenope, Isola C4, 80143 Napoli (Italy)

Phone: +390815476723.

# 1. Introduction

The two-dimensional Shallow water Equations (SWE) are a very popular mathematical model for flooding simulations in urbanized and non-urbanized areas, and a variety of numerical methods are now available to tackle their approximate solution. Among the others, Finite Volume schemes with interface Riemann solvers (Toro 2001) have been successful for their ability to manage flow discontinuities such as shocks and wet-dry frontiers. In addition, smart numerical approaches in Finite Volume schemes allow satisfying properties such as depth positivity, well-balancing, non-linear stability (Bouchut 2004).

Despite their success, these methods are computationally expensive, and researchers are currently making efforts along different research directions to speed up computations. Among the others, it is possible to cite research fields such as parallel computing (Delis and Mathioudakis 2009), Porous Shallow water Equations in urbanized areas (Varra et al. 2020), Finite Difference schemes avoiding the approximation of a Riemann problem (Stelling and Duinmeijer 2003), and the adoption of simplified forms of the SWE model such as the Local Inertia Approximation (LInA, Bates et al. 2010) and the Zero-Inertia approximation (Xanthopoulos and Koutitas 1976).

Simplified versions of the SWE model are obtained by canceling the equation terms that are negligible with respect to the others in appropriate conditions. For example, the LInA model arises from cancelation of the convective acceleration terms (de Almeida and Bates 2013), while the Zero-Inertia model is obtained by neglecting both local and convective accelerations (Xanthopoulos and Koutitas 1976). However, the friction forces acting at the flow bed are important in most applications for physical credibility of the models, and for this reason they are always retained. In two-dimensional flooding models, the friction forces per unit of weight are represented by the dimensionless friction vector $\mathbf{S}_f = \begin{pmatrix} S_{f,x} & S_{f,y} \end{pmatrix}^T$, where $S_{f,x}$ and $S_{f,y}$ are the component of $\mathbf{S}_f$ along the horizontal coordinate axes $x$ and $y$, respectively, while $T$ is the matrix transpose symbol. If bed roughness is isotropic, the friction vector is parallel to the flow velocity, implying that $S_{f,x}$ and $S_{f,y}$ are coupled because they both depend on the components of the velocity along *x* and *y* (Viero and Valipour 2017).

Due to the availability of raster digital elevation models as topographic input, the LInA and the Zero-Inertia models are often solved on square staggered grids (Bates and De Roo 2000). The two horizontal components of velocity are evaluated in different points of the grid, implying that the evaluation of friction is accomplished by means of some form of averaging on a large computational stencil (Hromadka and Chintu Lai 1985, Galbiati and Savi 1995, Wasantha Lal 1998, Bradbrook et al. 2004, Panday and Huyakorn 2004, Yu and Lane 2006, de Almeida and Bates 2013, Leandro et al. 2014).

Alternative to the averaging, acceleration of staggered grid algorithms can be obtained by decoupling the components of $\mathbf{S}_f$, i.e. computing $S_{f,x}$ and $S_{f,y}$ by separately using the *x* and *y* components of the velocity. Until now, the decoupled friction approach has been favored by many Authors (Bates and De Roo 2000, Horritt and Bates 2001, 2002, Hunter et al. 2005, 2007, Fewtrell et al. 2008, Dottori and Todini 2011, Neal et al. 2011, 2012, 2018, de Almeida et al. 2012, Falter et al. 2013, Yamazaki et al. 2013, Coulthard et al. 2013, Mendicino et al. 2015, Savage et al. 2016, Nguyen et al. 2016, Mateo et al. 2017, Sridharan et al. 2020) because it is easily programmed, more computationally efficient (Hunter et al. 2007), and numerical results have been found satisfactory when the topographic control is dominant (Bates and De Roo 2000, Horritt and Bates 2001). Nonetheless, it has been recognized that the decoupled-friction approach does not faithfully reflect the structure of the models derived from first principles (Hunter et al. 2007), and issues arise in urbanized areas (Hunter et al. 2007). It is believed that the calibration of friction coefficients may reduce the discrepancies between coupled- and decoupled-friction models (Horritt and Bates 2001), but the question if the errors introduced in decoupled-friction models can be compensated by calculations speed-up remains still open.

Due to the lack of consensus about the viability of decoupled-friction approach, in the present paper the question is tackled from the theoretical and numerical point of view. In particular, the coupled and decoupled definitions of the friction vector are compared, and the influence of these definitions on the mathematical structure of flooding models is explored, demonstrating that the decoupled friction definition causes the loss of rotational invariance of the models. This theoretical finding allows the interpretation of the numerical results, which confirm the lack of rotational invariance of decoupled-friction models. The paper is organized as follows: in Section 2, the friction vector is considered from a theoretical point of view; in Section 3, the LInA and the Zero-Inertia models with coupled and decoupled friction are presented, and four numerical schemes from the literature are used to solve a case study with radial symmetry; the theoretical and numerical findings are discussed in Section 4 while, finally, conclusions are drawn in Section 5.

**2. Analysis of the friction term**

If the friction force is parallel to the flow velocity, it is possible to write

(1) $S_{f,x} = \dfrac{q_x}{\|\mathbf{q}\|}\|\mathbf{S}_f\|, \quad S_{f,y} = \dfrac{q_y}{\|\mathbf{q}\|}\|\mathbf{S}_f\|.$

In Eq. (1), the meaning of the symbols is as follows: $q_x = hu$ and $q_y = hv$ are the components along $x$ and $y$, respectively, of the unit-width discharge vector $\mathbf{q} = \begin{pmatrix} q_x & q_y \end{pmatrix}^T$; $h$ is the flow depth; $u$ and $v$ are the component along $x$ and $y$, respectively, of the depth-averaged flow velocity; $\|\mathbf{S}_f\| = \sqrt{S_{f,x}^2 + S_{f,x}^2}$ is the *modulus* of $\mathbf{S}_f$ and $\|\mathbf{q}\| = \sqrt{q_x^2 + q_y^2}$ is the *modulus* of $\mathbf{q}$.

Despite its empirical nature, the Manning's formula has been justified theoretically (Gioia and Bombardelli 2002) and it is valid in a large variety of circumstances, including vegetated beds with high submergence (Gualtieri et al. 2018). For this reason, the expression

$$(2) \quad \|\mathbf{S}_f\| = n_M^2 \|\mathbf{q}\|^2 / h^{10/3} = n_M^2 (q_x^2 + q_y^2) / h^{10/3}$$

is usually adopted for the evaluation of the friction vector modulus. Eq. (2), where $n_M$ is the Manning's friction coefficient, implies that the components of the friction vector are

$$(3) \quad S_{f,x} = \frac{n_M^2 q_x \sqrt{q_x^2 + q_y^2}}{h^{10/3}}, \quad S_{f,y} = \frac{n_M^2 q_y \sqrt{q_x^2 + q_y^2}}{h^{10/3}}.$$

The inspection of Eq. (3) clearly shows that the components of the friction vector are coupled because they simultaneously depend on $q_x$ and $q_y$.

In the literature of simplified flooding models, it has become popular the definition of a decoupled friction vector $\mathbf{S'}_f = \left( S'_{f,x} \quad S'_{f,y} \right)^T$, where the components are calculated separately in a strictly one-dimensional fashion. In the case of the Manning's formula, the corresponding components are

$$(4) \quad S'_{f,x} = \frac{n_M^2 q_x |q_x|}{h^{10/3}}, \quad S'_{f,y} = \frac{n_M^2 q_y |q_y|}{h^{10/3}}.$$

Notably, the decoupled friction vector $\mathbf{S'}_f$ is not parallel to the flow vector $\mathbf{q}$, because $S'_{f,x}$ is not proportional to $q_x$ but to $q_x |q_x|$ and $S'_{f,y}$ is not proportional to $q_y$ but to $q_y |q_y|$.

In the present section, the coupled and decoupled definitions of the friction vector will be compared, and the influence of these definitions on the mathematical structure of flooding models will be explored.

**2.1 Modulus and angle of the decoupled-friction vector**

The *modulus* of the decoupled friction vector is

$$(5) \quad \|\mathbf{S'}_f\| = \sqrt{S'^2_{f,x} + S'^2_{f,x}} = \sqrt{\frac{n_M^4 q_x^4}{h^{20/3}} + \frac{n_M^4 q_y^4}{h^{20/3}}} = \frac{n_M^2 \sqrt{q_x^4 + q_y^4}}{h^{10/3}},$$

which differs from that of coupled friction vector in Eq. (2). If $\alpha$ is the angle between the friction vector $\mathbf{S}_f$ and the $x$ axis, the corresponding tangent is $\tan\alpha = S_{f,y}/S_{f,x} = q_y/q_x$. This allows calculating the ratio $\|\mathbf{S'}_f\|/\|\mathbf{S}_f\|$ between the *moduli* of the vectors $\mathbf{S'}_f$ and $\mathbf{S}_f$ as

$$(6) \quad \frac{\|\mathbf{S'}_f\|}{\|\mathbf{S}_f\|} = \frac{\sqrt{q_x^4 + q_y^4}}{q_x^2 + q_y^2} = \frac{\sqrt{q_x^4 + q_x^4 \tan^4\alpha}}{q_x^2 + q_x^2 \tan^2\alpha} = \frac{\sqrt{1 + \tan^4\alpha}}{1 + \tan^2\alpha}.$$

The ratio $\|\mathbf{S'}_f\|/\|\mathbf{S}_f\|$ ranges in the interval $[1/\sqrt{2}, 1]$ for $\alpha \in [0, 2\pi]$ (see Figure 1a), and this means that the friction force calculated by means of the decoupled formulas is systematically smaller than the coupled friction vector forces. The maximum error is attained for angles $\alpha = (2n+1)\pi/4$ with $n = 0, 1, 2, 3$, i.e. when the vector $\mathbf{S}_f$ is aligned with the bisectors of the reference framework quadrants. The average value of $\|\mathbf{S'}_f\|/\|\mathbf{S}_f\|$ is

$$(7) \quad E\left[\frac{\|\mathbf{S'}_f\|}{\|\mathbf{S}_f\|}\right] = \frac{1}{2\pi} \int_0^{2\pi} \frac{\sqrt{\tan^4 \alpha + 1}}{\tan^2 \alpha + 1} d\alpha = 0.86,$$

showing that the average error between the vector *moduli* is of about 14%.

If $\beta$ is the angle between the friction vector $\mathbf{S'}_f$ and the *x* axis, the corresponding tangent is

$$(8) \quad \tan \beta = \frac{S'_{f,y}}{S'_{f,x}} = \frac{q_y |q_y|}{q_x |q_x|} = \tan \alpha |\tan \alpha|,$$

which differs from *tan* $\alpha$. This means that the decoupled friction vector is rotated with respect to the coupled friction vector by the angle $\beta$ - $\alpha$. The absolute deviation $|\beta - \alpha|$ ranges in the interval [0, 0.071$\pi$] (see Figure 1b), and the corresponding average value is

$$(9) \quad E\left[|\beta - \alpha|\right] = \frac{1}{2\pi} \int_0^{2\pi} \left|\arctan\left(\tan \alpha |\tan \alpha|\right) - \alpha\right| d\alpha = 0.046\pi,$$

which is small and corresponding to 8.28° degrees.

From this discussion, the apparent conclusions are that:

- the *modulus* of $\mathbf{S'}_f$ is smaller than the *modulus* of $\mathbf{S}_f$, and this may require some calibration of the friction coefficient;
- the vectors $\mathbf{S}_f$ and $\mathbf{S'}_f$, which coincide only for $\alpha = n\pi/2$ with $n$ = 0, 1, 2, 3, are quite similar in the average;
- the substitution of $\mathbf{S}_f$ with $\mathbf{S'}_f$ is safe.

These provisional conclusions seem to justify the use of the decoupled friction $\mathbf{S'}_f$ instead of the friction $\mathbf{S}_f$ in flooding mathematical and numerical models. In the next sub-section, it will be shown that the substitution of $\mathbf{S}_f$ with $\mathbf{S'}_f$ induces unexpected structural changes in flooding models that are strongly in contrast with the preceding observations.

[Insert Figure 1 about here]

**2.2 Rotation of the reference framework for the SWE model**

In Newton's mechanics, the galilean transformations (rotation of the reference framework by an angle $\theta$, translation with uniform velocity, shifting of the origin) preserve intervals of time and distance between simultaneous events (Arnold 1989). This implies that flooding models are congruent with Newton's mechanics only if their structure remains invariant under rotation of the reference framework.

While SWE simplifications such as LInA and Zero-Inertia models make often use of the decoupled friction vector, this never happens in the literature of SWE model itself. Nonetheless, the preservation of the rotational invariance for the SWE model only is considered here, because the corresponding LInA and Zero-Inertia simplifications trivially inherit this property. Under the assumption that Eqs. (1) and (2) are used for the friction vector definition , the SWE can be written as

$$(10) \quad \begin{aligned} &\frac{\partial h}{\partial t} + \frac{\partial q_x}{\partial x} + \frac{\partial q_y}{\partial y} = 0 \\ &\frac{\partial q_x}{\partial t} + \frac{\partial}{\partial x}\left(\frac{gh^2}{2} + \frac{q_x^2}{h}\right) + \frac{\partial}{\partial y}\frac{q_x q_y}{h} = -gh\frac{\partial z}{\partial x} - g\frac{n_M^2 q_x \|\mathbf{q}\|}{h^{7/3}}, \\ &\frac{\partial q_y}{\partial t} + \frac{\partial}{\partial x}\frac{q_x q_y}{h} + \frac{\partial}{\partial y}\left(\frac{gh^2}{2} + \frac{q_y^2}{h}\right) = -gh\frac{\partial z}{\partial y} - g\frac{n_M^2 q_y \|\mathbf{q}\|}{h^{7/3}} \end{aligned}$$

where $t$ is the time, $h$ is the flow depth, $z$ is the bed elevation, and $g = 9.81$ m/s$^2$ is the gravity acceleration, while all the other symbols have been already defined. The first of Eq. (10) represents the mass conservation, while the second and the third represent the momentum conservation along $x$ and $y$, respectively. In the following, the invariance of Eq. (10) to the frame reference rotation will be easily verified.

Let $OXY$ be a reference framework rotated with respect to the reference framework $Oxy$ by the angle $\theta$. The relation between the coordinates in the two reference frameworks is:

(11) $X = x\cos\theta + y\sin\theta, \quad Y = -x\sin\theta + y\cos\theta$.

If $\mathbf{Q} = (Q_X \quad Q_Y)^T$ is the unit-width discharge vector in the rotated framework, the relation between $\mathbf{Q}$ and $\mathbf{q}$ is

(12) $\begin{aligned} q_x &= Q_X \cos\theta - Q_Y \sin\theta \\ q_y &= Q_X \sin\theta + Q_Y \cos\theta \end{aligned}$.

From Eq. (12) it is evident that $\|\mathbf{q}\| = \sqrt{q_x^2 + q_y^2} = \sqrt{Q_X^2 + Q_Y^2} = \|\mathbf{Q}\|$, confirming that the *modulus* of the velocity vector remains unchanged under a rotation of the reference framework. Recalling that the relation between the derivatives in the two frameworks is

(13) $\begin{aligned} \frac{\partial}{\partial x} &= \cos\theta \frac{\partial}{\partial X} - \sin\theta \frac{\partial}{\partial Y} \\ \frac{\partial}{\partial y} &= \sin\theta \frac{\partial}{\partial X} + \cos\theta \frac{\partial}{\partial Y} \end{aligned}$,

the Eq. (10) can be rewritten in the rotated framework (see Appendix A) as:

$$(14)\quad\begin{aligned}&\frac{\partial h}{\partial t}+\frac{\partial Q_X}{\partial X}+\frac{\partial Q_Y}{\partial Y}=0\\&\frac{\partial Q_X}{\partial t}+\frac{\partial}{\partial X}\left(\frac{gh^2}{2}+\frac{Q_X^2}{h}\right)+\frac{\partial}{\partial Y}\frac{Q_XQ_Y}{h}=-gh\frac{\partial z}{\partial X}-g\frac{n_M^2Q_X\|\mathbf{Q}\|}{h^{7/3}}\\&\frac{\partial Q_Y}{\partial t}+\frac{\partial}{\partial X}\frac{Q_XQ_Y}{h}+\frac{\partial}{\partial Y}\left(\frac{gh^2}{2}+\frac{Q_Y^2}{h}\right)=-gh\frac{\partial z}{\partial Y}-g\frac{n_M^2Q_Y\|\mathbf{Q}\|}{h^{7/3}}\end{aligned}$$

The comparison between Eq. (10) and Eq. (14) shows that the structure of SWE model with coupled friction remains unchanged under the rotation framework, because the rotated systems of Eq (14) can be obtained from Eq. (10) by simply substituting $X$ to $x$, $Y$ to $y$, $Q_X$ to $q_x$, and $Q_Y$ to $q_y$. Interestingly, the reference framework rotation preserves the structure of the coupled friction terms because the modulus of the friction vector and its parallelism with the flow velocity vector are preserved. These findings imply that the computation results in the rotated framework are indistinguishable from the computation results in the original reference framework, congruently with the Newton's mechanics and with our physical intuition. In other words, all the reference frameworks are equally good for the simulation of floodings with coupled-friction models.

Is the preceding result applicable to the case of decoupled friction? The answer is no. With respect to the $Oxy$ reference framework, the SWE model with decoupled friction can be written as

$$(15)\quad\begin{aligned}&\frac{\partial h}{\partial t}+\frac{\partial q_x}{\partial x}+\frac{\partial q_y}{\partial y}=0\\&\frac{\partial q_x}{\partial t}+\frac{\partial}{\partial x}\left(\frac{gh^2}{2}+\frac{q_x^2}{h}\right)+\frac{\partial}{\partial y}\frac{q_xq_y}{h}=-gh\frac{\partial z}{\partial x}-g\frac{n_M^2q_x|q_x|}{h^{7/3}}\\&\frac{\partial q_y}{\partial t}+\frac{\partial}{\partial x}\frac{q_xq_y}{h}+\frac{\partial}{\partial y}\left(\frac{gh^2}{2}+\frac{q_y^2}{h}\right)=-gh\frac{\partial z}{\partial y}-g\frac{n_M^2q_y|q_y|}{h^{7/3}}\end{aligned}$$

The rotation of the reference framework by the angle $\theta$ supplies the rotated system

$$\frac{\partial h}{\partial t} + \frac{\partial Q_X}{\partial X} + \frac{\partial Q_Y}{\partial Y} = 0$$

(16)
$$\frac{\partial Q_X}{\partial t} + \frac{\partial}{\partial X}\left(\frac{gh^2}{2} + \frac{Q_X^2}{h}\right) + \frac{\partial}{\partial Y}\frac{Q_X Q_Y}{h} = -gh\frac{\partial z}{\partial X} - g\frac{n_M^2 f_1(\theta, Q_X, Q_Y)}{h^{7/3}},$$

$$\frac{\partial Q_Y}{\partial t} + \frac{\partial}{\partial X}\frac{Q_X Q_Y}{h} + \frac{\partial}{\partial Y}\left(\frac{gh^2}{2} + \frac{Q_Y^2}{h}\right) = -gh\frac{\partial z}{\partial Y} - g\frac{n_M^2 f_2(\theta, Q_X, Q_Y)}{h^{7/3}}$$

where the functions $f_1(\theta, Q_X, Q_Y)$ and $f_2(\theta, Q_X, Q_Y)$ are defined as

(17)
$$\begin{aligned}f_1(\theta, Q_X, Q_Y) &= \cos\theta(Q_X\cos\theta - Q_Y\sin\theta)|Q_X\cos\theta - Q_Y\sin\theta| \\ &+ \sin\theta(Q_X\sin\theta + Q_Y\cos\theta)|Q_X\sin\theta + Q_Y\cos\theta| \\ f_2(\theta, Q_X, V) &= -\sin\theta(Q_X\cos\theta - Q_Y\sin\theta)|Q_X\cos\theta - Q_Y\sin\theta| \\ &+ \cos\theta(Q_X\sin\theta + Q_Y\cos\theta)|Q_X\sin\theta + Q_Y\cos\theta|\end{aligned}.$$

The comparison between Eqs. (15) and (16) shows that the rotation of the reference framework causes the structural change of the decoupled-friction SWE model. In particular, the two components of the friction in the rotated framework are not proportional to $Q_X|Q_X|$ and $Q_Y|Q_Y|$ while Eq. (16) reduces to Eq. (15) only for $\theta = n\pi/2$ with n = 0, 1, 2, 3. The lack of rotational invariance implies that *the results of the SWE model with decoupled friction depend on the choice of the reference framework*. This characteristic is in contrast not only with Newton's mechanics but also with the physical intuition, and it represents a major loss of meaning of the model. In this case, it is evident that calculations are physically meaningful only if the flow is strictly one-dimensional and parallel to one of the coordinate axes.

**2.3 Rotation of the framework reference for LInA and Zero-Inertia models**

It is possible to demonstrate that the LInA and the Zero-Inertia models inherit the rotational invariance property of the SWE model. In particular, the LInA and the Zero-Inertia models with coupled friction terms preserve the rotational invariance, while the corresponding decoupled-friction models do not satisfy this fundamental property. This is easily proven by applying the procedure of Sub-section 2.2 to the simplified models with coupled and decoupled friction, but the complete demonstration is omitted here for the sake of brevity.

It follows that the results of LInA and Zero-Inertia models with coupled friction do not depend on the reference framework, while *the same models with decoupled friction suffer from dependency on the choice of the reference framework*. The computational aspects of this discouraging conclusion will be explored in the next Section.

## 3. Flooding computation with coupled/decoupled friction

In the present Section, the LInA and the Zero-Inertia mathematical models are presented and, for each friction version (coupled/decoupled) of these models, a numerical algorithm from the literature (Wasantha Lal 1998, Bates and De Roo 2000, de Almeida et al. 2012, de Almeida and Bates 2013) is briefly described. Finally, the results corresponding to a special case study are presented for each numerical model. For the sake of simplicity, details about the treatment of wetting-drying fronts are omitted because the case study refers to a fully wet flow field.

### 3.1 Mathematical and numerical models

In the present Sub-section, the mathematical and numerical LInA and Zero-Inertia models are presented. Despite the controversy about viability of the LInA model in flooding applications

(Cozzolino et al. 2019), this model is considered here because it is often applied after the decoupling of friction vector components.

### *3.1.1 LInA model with coupled friction*

The LInA model is obtained from the SWE model by neglecting the convective acceleration terms. Assuming that discontinuities are absent in the flow field, the LInA model with coupled friction is written as (de Almeida and Bates 2013)

$$(18) \quad \begin{aligned} &\frac{\partial h}{\partial t} + \frac{\partial q_x}{\partial x} + \frac{\partial q_y}{\partial y} = 0 \\ &\frac{\partial q_x}{\partial t} + gh\frac{\partial \eta}{\partial x} + g\frac{n_M^2 q_x \|\mathbf{q}\|}{h^{7/3}} = 0, \\ &\frac{\partial q_y}{\partial t} + gh\frac{\partial \eta}{\partial y} + g\frac{n_M^2 q_y \|\mathbf{q}\|}{h^{7/3}} = 0 \end{aligned}$$

where $\eta = h + z$ is the free-surface elevation.

For the approximate solution of Eq. (18), the numerical scheme on staggered grid by de Almeida and Bates (2013) is considered here. After that the physical domain is subdivided in square cells with sides of length $\Delta s$ that are aligned with the $x$ and $y$ axes, the numerical approximation of $h$ is collocated at the center of the cell (black dots in Figure 2). Correspondingly, the numerical approximations of $q_x$ and $q_y$ are collocated at the center of interfaces which are normal to the axis $x$ (grey squares in Figure 2) and to the axis $y$ (grey triangles), respectively.

The component $q_x$ of the unit-width discharge at the interface ($i$-1/2,$j$) between the cells ($i$-1,$j$) and ($i$,$j$) is updated by means of

$$(19) \quad q_{x,i-1/2,j}^{n+1} = \frac{\lambda q_{x,i-1/2,j}^n + 0.5(1-\lambda)\left(q_{x,i+1/2,j}^n + q_{x,i-3/2,j}^n\right) - gh_{i-1/2,j}^n \Delta t \left(\eta_{i,j}^n - \eta_{i-1,j}^n\right)/\Delta s}{1 + g\Delta t n_M^2 \|\mathbf{q}_{i-1/2,j}^n\| / \left(h_{i-1/2,j}^n\right)^{7/3}}.$$

In Eq. (19), the meaning of the symbols is as follows: $q_{x,i-1/2,j}^n$ is a numerical approximation of $q_x$ at the interface ($i$-1/2,$j$) at the time level $t_n$; $\lambda$ is a relaxation factor chosen in the interval [0, 1]; $\Delta t = t_{n+1} - t_n$ is the time step; $\eta_{i,j}^n = h_{i,j}^n + z_{i,j}$ is a numerical approximation of the free-surface elevation in the cell ($i$,$j$), where $h_{i,j}^n$ and $z_{i,j}$ are the cell-averaged flow depth and bed elevation, respectively; $h_{i-1/2,j}^n$ is a numerical approximation of $h$ at the interface ($i$-1/2,$j$); finally, $\|\mathbf{q}_{i-1/2,j}^n\| = \sqrt{\left(q_{x,i-1/2,j}^n\right)^2 + \left(q_{y,i-1/2,j}^n\right)^2}$ is an approximation of $\|\mathbf{q}\|$ at the interface ($i$-1/2,$j$).

Notably, the numerical counterparts of $h$ and $q_y$ are not defined at the interface ($i$-1/2,$j$), which is parallel to the $y$ axis. For this reason, the approximations

(20) $h_{i-1/2,j}^n = \max\left(\eta_{i-1,j}^n, \eta_{i,j}^n\right) - \max\left(z_{i-1,j}, z_{i,j}\right)$

and the average

(21) $q_{y,i-1/2,j}^n = 0.25\left(q_{y,i-1,j-1/2}^n + q_{y,i-1,j+1/2}^n + q_{y,i,j-1/2}^n + q_{y,i,j+1/2}^n\right)$

are used.

A similar formulation, with obvious changes, is used for updating the unit-width discharge $q_y$ at cell interfaces. Once that the discharges are updated, it is possible to update the cell-averaged flow depth $h_{i,j}^n$ by means of

(22) $h_{i,j}^{n+1} = h_{i,j}^n - \dfrac{\Delta t}{\Delta s}\left(q_{x,i+1/2,j}^{n+1} - q_{x,i-1/2,j}^{n+1}\right) - \dfrac{\Delta t}{\Delta s}\left(q_{y,i,j+1/2}^{n+1} - q_{y,i,j-1/2}^{n+1}\right).$

[Insert Figure 2 about here]

Note that the calculation of $q_{x,i-1/2,j}^{n+1}$ involves only $q_{x,i-1/2,j}^{n}$ when $\lambda = 1$, while $q_{x,i-3/2,j}^{n}$ and $q_{x,i+1/2,j}^{n}$ are also involved when $\lambda < 1$. In other words, the numerical stencil for the calculation of $q_{x,i-1/2,j}^{n+1}$ is expanded in the $x$ direction when $\lambda < 1$.

### 3.1.2 LInA model with decoupled friction

The LInA model with decoupled friction can be written as (de Almeida et al. 2012)

$$(23) \quad \begin{aligned} &\frac{\partial h}{\partial t} + \frac{\partial q_x}{\partial x} + \frac{\partial q_y}{\partial y} = 0 \\ &\frac{\partial q_x}{\partial t} + gh\frac{\partial \eta}{\partial x} + g\frac{n_M^2 q_x |q_x|}{h^{7/3}} = 0 \\ &\frac{\partial q_y}{\partial t} + gh\frac{\partial \eta}{\partial y} + g\frac{n_M^2 q_y |q_y|}{h^{7/3}} = 0 \end{aligned}$$

The numerical scheme for its solution (de Almeida et al. 2012) is identical to the scheme used for the case of coupled friction, but now $q_x$ is updated without considering the contribution of $q_y$ in the friction calculation, and $q_y$ is updated neglecting the contribution of $q_x$. For this reason, Eq. (19) is substituted by

$$(24) \quad q_{x,i-1/2,j}^{n+1} = \frac{\lambda q_{x,i-1/2,j}^{n} + 0.5(1-\lambda)\left(q_{x,i+1/2,j}^{n} + q_{x,i-3/2,j}^{n}\right) - gh_{i-1/2,j}^{n}\Delta t\left(\eta_{i,j}^{n} - \eta_{i-1,j}^{n}\right)/\Delta s}{1 + g\Delta t n_M^2 |q_{x,i-1/2,j}^{n}|/\left(h_{i-1/2,j}^{n}\right)^{7/3}},$$

while Eq. (21) is not used. Due to the smaller number of operations, the algorithm with decoupled friction is faster than the algorithm with coupled friction.

### 3.1.3 Zero-Inertia model with coupled friction

The Zero-Inertia model is obtained from the SWE model by neglecting both the local and the convective acceleration terms. The model version with coupled friction can be written as (Xanthopoulos and Koutitas 1976)

$$(25) \quad \begin{aligned} &\frac{\partial h}{\partial t} + \frac{\partial q_x}{\partial x} + \frac{\partial q_y}{\partial y} = 0 \\ &\frac{\partial \eta}{\partial x} + \frac{n_M^2 q_x \|\mathbf{q}\|}{h^{10/3}} = 0 \\ &\frac{\partial \eta}{\partial y} + \frac{n_M^2 q_y \|\mathbf{q}\|}{h^{10/3}} = 0 \end{aligned}.$$

For computational purposes, it is convenient to rewrite the system in the form (Hromadka and Chintu Lai 1985)

$$(26) \quad \begin{aligned} &\frac{\partial h}{\partial t} + \frac{\partial q_x}{\partial x} + \frac{\partial q_y}{\partial y} = 0 \\ &q_x = -\frac{h^{5/3}}{n_M \sqrt{\|\mathbf{S}\|}} \frac{\partial \eta}{\partial x} \\ &q_y = -\frac{h^{5/3}}{n_M \sqrt{\|\mathbf{S}\|}} \frac{\partial \eta}{\partial y} \end{aligned}$$

where $\mathbf{S} = (\partial \eta/\partial x \quad \partial \eta/\partial y)^T$ is the free-surface gradient and $\|\mathbf{S}\| = \sqrt{(\partial \eta/\partial x)^2 + (\partial \eta/\partial y)^2}$ is the free-surface gradient *modulus*.

Following Wasantha Lal (1998), the component $q_x$ of the unit-width discharge at the interface ($i$-1/2,$j$) between the cells ($i$-1,$j$) and ($i$,$j$) at the time level $t_n$ is calculated by means of

$$(27) \quad q_{x,i-1/2,j}^n = \frac{\left(h_{i-1/2,j}^n\right)^{5/3}}{n_M \sqrt{\|\mathbf{S}_{i-1/2,j}^n\|}} \frac{\eta_{i,j}^n - \eta_{i-1,j}^n}{\Delta s},$$

where

$$(28) \quad \|\mathbf{S}_{i-1/2,j}^n\| = \sqrt{\left(\frac{\eta_{i,j}^n - \eta_{i-1,j}^n}{\Delta s}\right)^2 + \left(\frac{1}{2}\frac{\eta_{i,j+1}^n - \eta_{i,j-1}^n}{2\Delta s} + \frac{1}{2}\frac{\eta_{i-1,j+1}^n - \eta_{i-1,j-1}^n}{2\Delta s}\right)^2}$$

is a numerical approximation of $\|\mathbf{S}\|$ at the interface $(i-1/2,j)$, while

$$(29) \quad h_{i-1/2,j}^n = \frac{h_{i-1,j}^n + h_{i,j}^n}{2}$$

is a numerical approximation of $h$ at the same interface. Similar formulas are used for the calculation of the unit-width discharges on the other interfaces, allowing the update of $h$ in the cell $(i,j)$ by means of the explicit formula

$$(29) \quad h_{i,j}^{n+1} = h_{i,j}^n - \frac{\Delta t}{\Delta s}\left(q_{x,i+1/2,j}^n - q_{x,i-1/2,j}^n\right) - \frac{\Delta t}{\Delta s}\left(q_{y,i,j+1/2}^n - q_{y,i,j-1/2}^n\right).$$

*3.1.4 Zero-Inertia model with decoupled friction*

The Zero-Inertia model with decoupled friction can be written as

$$\frac{\partial h}{\partial t} + \frac{\partial q_x}{\partial x} + \frac{\partial q_y}{\partial y} = 0$$

(30) $\quad \dfrac{\partial \eta}{\partial x} + \dfrac{n_M^2 q_x |q_x|}{h^{10/3}} = 0$ ,

$$\frac{\partial \eta}{\partial y} + \frac{n_M^2 q_y |q_y|}{h^{10/3}} = 0$$

which can be rewritten as (Hunter et al. 2005)

$$\frac{\partial h}{\partial t} + \frac{\partial q_x}{\partial x} + \frac{\partial q_y}{\partial y} = 0$$

(31) $\quad q_x = -\dfrac{h^{5/3}}{n_M} \mathrm{sgn}\left(\dfrac{\partial \eta}{\partial x}\right) \sqrt{\left|\dfrac{\partial \eta}{\partial x}\right|}$ ,

$$q_y = -\frac{h^{5/3}}{n_M} \mathrm{sgn}\left(\frac{\partial \eta}{\partial y}\right) \sqrt{\left|\frac{\partial \eta}{\partial y}\right|}$$

where *sgn*(•) represents the *signum* function. Following Bates and De Roo (2000), the component $q_x$ of the unit-width discharge at the interface (*i-1/2,j*) between the cells (*i-1,j*) and (*i,j*) at the time level $t_n$ is calculated by means of

(31) $\quad q_{x,i-1/2,j}^n = -\dfrac{\left(h_{i-1/2,j}^n\right)^{5/3}}{n_M} \mathrm{sgn}\left(\dfrac{\eta_{i,j}^n - \eta_{i-1,j}^n}{\Delta s}\right) \sqrt{\left|\dfrac{\eta_{i,j}^n - \eta_{i-1,j}^n}{\Delta s}\right|}$ ,

where Eq. (20) is used to approximate the flow depth $h_{i-1/2,j}^n$ at the interface (*i-1/2,j*). Similar formulas, with obvious changes, are used for the calculation of the unit-width discharges on the interfaces with normal *y*, allowing the update of *h* in the cell (*i,j*) by means of Eq. (29).

The numerical scheme with decoupled friction is faster than the scheme with coupled friction because the approximation of the complete free-surface gradient at cell interfaces is avoided.

**3.2 Numerical case study**

The mathematical and numerical models of Sub-section 3.1 are evaluated considering a case study inspired by that introduced in Wasantha Lal (1998). The physical domain consists of a square area with side $L$ = 160930 m characterised by uniform bed elevation ($z$ = 0 m) and uniform Manning's roughness coefficient ($n_M$ = 0.03 s m$^{-1/3}$). The origin $O$ of the reference framework $Oxy$ is at the centre of the domain, with axes $x$ and $y$ parallel to the square sides. The initial conditions are characterised by sinusoidal free-surface elevation

$$(32) \quad \eta(x,y,0) = \begin{cases} \left[0.4575 + 0.1525 \cos \dfrac{\pi\sqrt{x^2+y^2}}{32187.9}\right] \text{m}, & \sqrt{x^2+y^2} \leq 32187.9 \text{ m} \\ 0.305 \text{ m}, & \sqrt{x^2+y^2} > 32187.9 \text{ m} \end{cases},$$

while wall boundary conditions are considered at the limits of the domain.

The expected solution of the problem remains radially symmetric with time and the fronts of the propagating wave are expected to keep a circular shape for $t > 0$, because the initial conditions are radially symmetric while roughness and bed elevation are uniform.

Due to the symmetry of initial conditions with respect to the $x$ and $y$ axes, it is sufficient to consider the square numerical domain with side $l$ = $L/2$ = 80465 m for $x \geq 0$ and $y \geq 0$, with origin $O$ of the reference framework located at the lower left vertex of the domain (see for example Figure 3). The computer programs for the numerical models compared here are written in Microsoft Visual

Basic 6.0, then compiled and run on a laptop computer with Vista Home Basic operative system, Intel T2080 processor (1.73 GHz clock), and 2.00 Gb RAM.

### *3.2.1 Zero-Inertia model with coupled and decoupled friction*

The results of the Zero-Inertia model with coupled friction (numerical scheme of sub-section 3.1.3) are presented at time $t = 60000$ s in the left column of Figure 3 for three different densities of the numerical grid. The corresponding uniform grid spacing and constant time steps used in the computations are detailed in Table 1. The free-surface elevation for the coarse grid computations is represented in the upper panel (Figure 3a). The inspection of this panel shows that, congruently with expectation, the free-surface elevation propagates symmetrically along all the directions originating from $O$, exhibiting good isotropy properties. The grid refinement in Figure 3c (medium grid) and Figure 3e (refined grid) confirms the radial symmetry of the solution, congruently with the physical intuition.

The results of the corresponding decoupled friction model (numerical scheme of sub-section 3.1.4) are presented at time $t = 60000$ s in the right column of Figure 3. A closer inspection of Figure 3b shows that the effect of friction decoupling is the distortion of the propagating wave, which progressively tends to a sort of smoothed square, because the free-surface contour lines tend to be parallel to the coordinate axes. This implies that a sort of preferential line of wave propagation arises along the bisector of the numerical domain, where the flow is faster. Also in the case of Zero-Inertia model with decoupled friction, the distorted solution does not change significantly with the refinement of the grid in Figure 3d and 3f, meaning that the structural issue of the decoupled-friction model cannot be solved by refining the grid.

By convention, we set the limit of the expanding wave at the free-surface elevation $\eta = 0.31$ m, which is also the contour line with lower value plotted in Figure 3. The computations show that the area with $\eta \geq 0.31$ m supplied by the coupled friction model for the refined grid is $2.11 \cdot 10^3$ km$^2$, while the corresponding area supplied by the decoupled friction model is $2.25 \cdot 10^3$ km$^2$. The increase of the area occupied by the expanding wave is of about 6.2 %, implying that the decoupled-friction wave is faster than the coupled-friction one.

[Insert Figure 3 about here]

[Insert Table 1 about here]

*3.2.2 LInA model with coupled and decoupled friction*

The LInA model with coupled friction (numerical scheme of sub-section 3.1.1) is run with relaxation parameter $\lambda = 1.0$, using the time step and grid spacing of Table 1 to allow a fair comparison with the Zero-Inertia model. The initial velocity must be specified, and its value has been set to zero everywhere for the sake of simplicity. The corresponding results at time $t = 60000$ s are presented in the left column of Figure 4. The inspection of Figure 4a (coarse grid), Figure 4c (medium grid), and Figure 4e (refined grid), confirms that the LInA model with coupled friction satisfies the radial symmetry of the problem, similarly to the case of the Zero-Inertia model with coupled friction.

The calculations are repeated with the decoupled-friction LInA model (numerical scheme of sub-section 3.1.2) with relaxation parameter $\lambda = 1.0$, and the corresponding results are represented in the right column of Figure 4 for the grids of Table 1. Again, the use of a decoupled-friction approach distorts the propagating wave fronts, which assume a smoothed square shape, while a preferential propagation line is individuated along the bisector of the numerical domain. In conclusion, the LInA model with decoupled friction exhibits the same anisotropy phenomenon of the Zero-Inertia model with decoupled friction, and this structural issue cannot be solved by means of grid refinement.

The area with $\eta \geq 0.31$ m supplied by the coupled-friction model for the refined grid is $2.14 \cdot 10^3$ km$^2$, while the corresponding area supplied by the decoupled friction model is $2.26 \cdot 10^3$ km². Also for the LInA model, the decoupled-friction wave is faster than the coupled-friction one, because the increase of the area occupied by the expanding wave is of about 5.6%.

[Insert Figure 4 about here]

*3.2.3 Run times*

The use of a constant time step allows the comparison of the algorithm run times, which are reported in Table 2 for the computations on the medium size grid. It is evident that these run times are only indicative, and may vary depending on the case study at hand, on the optimization of the algorithm construction, on the programming language, and on the hardware used.

The inspection of Table 2 first shows that, for given constant time step, the LInA model is computationally more expensive than the Zero-Inertia model. This is expected because the quantities $h$, $q_x$, and $q_y$, are advanced in time in the LInA model, while only $h$ evolves in the Zero-Inertia model. The comparison between coupled and decoupled models show that the Zero-Inertia model computations with decoupled friction allows a 1.25 times speedup over the corresponding coupled-friction computations, while the LInA model with decoupled friction allows a 1.17 times speedup over the LInA model with coupled friction.

It is possible to compare these results with the speedups obtained in the literature by means of parallel computing. Leandro et al. (2014) obtained up to a 1.7 times speedup for the Zero-Inertia model on 12 parallel cores, while Artichowicz and Gaşiorowski (19) obtained up to 4.86 times speedup on a 8 cores architecture. Similarly, Neal et al. (2018) reported up to 60 times speed up for an optimized version of the LInA model on 16 cores. It is evident the speedup obtained by decoupling the friction vector is modest with respect to the speed up obtained by parallel computing.

[Insert Table 2 about here]

## 4. Discussion

In the present Section, the theoretical results of Section 2 are discussed considering the numerical results of Section 3. An additional remark is made about solution distortion caused by the decoupling introduced by numerics.

### 4.1 A commentary on friction decoupling

It is commonly believed (Horritt and Bates 2001) that the Manning's friction coefficient can be adjusted to improve the results of decoupled-friction models. In order to understand the limits of this procedure, a calibration exercise has been accomplished by searching the value of $n_M$ that makes the decoupled-friction LInA model supply a moving wave area equal to the one supplied by the coupled-friction model at time $t = 60000$ s on the refined grid. After some attempt, this value has been found to be $n_M = 0.0325$ s m$^{-1/3}$, which differs by 8% with the original value $n_M = 0.03$ s m$^{-1/3}$. The results are compared in Figure 5, where the contour corresponding to $\eta = 0.31$ m for the coupled-friction model is plotted with a thin black line, while the same contour for the decoupled-friction model is plotted with a thick dashed line. The inspection of the figure clearly shows that, despite the two curves enclose the same area, the shape of the decoupled-friction moving wave remains different from the one of the coupled-friction wave. In other words, friction calibration cannot reduce the systematic difference between wave propagation celerities on $x/y$ axes and quadrant bisector.

[Insert Figure 5 about here]

The influence of the reference framework rotation on the model results is not an academic question and it has practical relevance. Alignment of the rectangular grid (i.e. of the reference framework) with the main direction of flow propagation is a common modelling practice (Bradbrook et al. 2004), because this procedure simplifies the setting of inflow and outflow boundary conditions. The connection between the grid alignment and the loss of rotational invariance studied in Sub-section 2.2 can be elucidated by considering Figure 6. In the left panel (Figure 6a), the solution of the LInA model with decoupled friction on refined grid is represented considering the reference framework $Oxy$ where the $x$ axis is aligned along the direction West-East (W-E) and the $y$ axis is aligned along the direction South-North (S-N). In this case, the stretching direction of the solution (dashed line) is parallel to the direction from South-West to North-East. Consider now the reference framework $OXY$ (Figure 6b), which is rotated by the angle $\theta = \pi/4$ with respect to the reference framework $Oxy$. In this case, the computations are characterised by stretching direction aligned with the South-North direction, confirming that the loss of rotational invariance causes the dependence of the solution on the alignment of the reference framework.

At this point, the model user may ask which is the correct computational grid alignment, and the answer is that all the grid alignments are equally wrong in decoupled-friction models. Flooding models, which incorporate a certain amount of theories, aim not only to computation but also to explanation (Clark et al. 2016). In particular, it is expected not only that a flooding model supplies acceptable results that are validated by available field data, but also that it supplies correct results for the correct reason (congruence with underlying physical laws). Satisfaction of this requirement supports the credibility of the model also in circumstances that are different from those that were chosen for validation. Unfortunately, the findings above are particularly discouraging because the friction decoupling introduces a loss of physical meaning that is hardly recognizable in topography-dominated applications (Bates and De Roo 2000, Horritt and Bates 2001) but it becomes evident in flat areas with truly two-dimensional applications. From the modeller's point of view, the decrease of friction forces and the dependence on the reference framework may be acknowledged, but they are hardly quantified *a priori* because comparison with a coupled-friction model is needed in each practical application. On the other hand, the common user may be unaware of an error that increases the model epistemic uncertainty (Oberkampf et al. 2002). It is evident that the conceptual and computational error, which in the writers' opinion is not compensated by the modest speed-up of calculations, can be eliminated by adopting a coupled-friction model.

It must be noted that these adverse effects did not emerge in the literature of decoupled-friction flooding models until now, and this can be easily justified considering that decoupled-friction flooding models have been often validated first by means of some purely one-dimensional case study in controlled conditions and then by comparing the model results with field data available from historical floods (de Almeida et al. 2012, Sridharan et al. 2020). Of course, one-dimensional case studies are not able to shed light on two-dimensional modelling intricacies, while real-world case studies introduce interpretation difficulties originated by complicated topographies and/or land cover characteristics. This explains why a truly two-dimensional flow propagation problem characterized by uniform bed elevation and uniform friction coefficient was chosen in Section 3 to evaluate the effects introduced by friction vector decoupling .

[Insert Figure 6 about here]

**4.2 A remark on the equivalence by Cellular Automata and Finite Volume schemes**

In the recent past, a class of algorithms for flooding simulation termed Cellular Automata (Thomas and Nicholas 2002, Rinaldi et al. 2007, Ghimire et al. 2013, Liu et al. 2015) has emerged for flood delineation applications. In these algorithms, the flow field is represented by means of a two-dimensional lattice, and the evolution of the flow storage in each lattice cell is driven by the state of the surrounding elements and by an appropriate rule of interaction between cells (Cai et al. 2014).

The rule for the evolution of the cell storage is often simple and empirical but, in many cases, a more physical approach can be used to calculate flows between cells.

An important adverse characteristic of Cellular Automaton algorithms with simple evolution rules is the fact that their results are strongly affected by the geometry of the underlying grid (Schönfisch 1995), because plane front waves appear where isotropic propagation is expected. For example, the binary spread on a rectangular grid with Neumann neighbourhood produces a diamond shape front while structured triangular grids produce a hexagonal shape (Ortigoza 2015). This means that Cellular Automata with simple rules are anisotropic at local scale, i.e. the signals propagate with different celerities in different directions, and this anisotropy reverberates at global scale (Schönfisch 1997) producing spurious geometric regularities. These effects can be quantified, and Weimar et al. (1992) found that the ratio between wave celerities in diagonal and horizontal directions is $1/\sqrt{2}$ in Cellular Automata algorithms with square lattice and simple evolution rules.

The continuous-state Cellular Automata paradigm (discrete variables in space and a rule for their evolution) is generic enough to embrace also classic numerical schemes for the solution of partial differential equations such as Finite Difference (Yang and Young 2006) and Finite Volume schemes. Not surprisingly, Caviedes-Voullième et al. (2018) have correctly stated that many Cellular Automata algorithms for flooding delineation correspond to Finite Volume discretizations of an appropriate system of differential equations. Nonetheless, these Authors have failed in individuating the correct correspondence between algorithms and differential equations. For example, the discrete model by Bates and De Roo (2000), which can be viewed as a Cellular Automata algorithm where the fluxes between cells are calculated by means of the one-dimensional Zero-Inertia model, is not a Finite Volume discretization of Eq. (25) as stated by Caviedes-Voullième et al. (2018), but it is a discretization of Eq. (30).

Of course, there is a formal correspondence between flow differential models with decoupled friction and Cellular Automata with simplified exchange rules between cells. Not surprisingly, Figures 3 and 4 (right columns) show that decoupled-friction models produce propagating waves whose isolines tend to be parallel to the grid axes. Strikingly, the $1/\sqrt{2}$ ratio between wave celerities in diagonal and horizontal directions found by Weimar et al. (1992) coincides with the ratio $\|\mathbf{S'}_f\|/\|\mathbf{S}_f\|$ for the bisectors of the quadrants in the flow propagation models with decoupled friction considered here.

**4.3 Decoupling introduced by numerical methods**

In the present paper, it has been demonstrated that the decoupling of the friction vector introduces the loss of the mathematical model rotational invariance and the distortion of numerical solutions. In addition to these findings, we want to show how numerical choices may introduce unwanted decoupling effects even in the case that the original mathematical model is based on coupled friction.

In Section 3, the discretization of the LInA numerical model by de Almeida and Bates (2013) has been constantly applied with relaxation parameter $\lambda = 1$, which ensures isotropy of the numerical model in the case of isotropic treatment of the friction. Nonetheless, fixed values of $\lambda$ in the range [0.7, 0.9] (de Almeida et al. 2012, de Almeida and Bates 2013) or dynamically variable values with $\lambda \leq 1$ (Sridharan et al. 2020) are recommended in the literature in order to ensure stability of computations in the case of flow propagation on dry bed with small friction. For this reason, the coupled-friction LInA model of Sub-section 3.1.1 with $\lambda = 1$ is contrasted in Figure 7 (left column) with the same model where the parameter $\lambda = 0.7$ is applied (right column).

The inspection of the figure shows that the coupled-friction by de Almeida and Bates (2013) with $\lambda < 1$ causes the distortion of the propagating front, which is progressively converted into a straight line inclined by $-\pi/4$ (right column of Figure 7). This effect is not present in the same model with $\lambda = 1$ (left column of Figure 7). It is evident that the expansion of the $q_x$ stencil along $x$ and the $q_y$ stencil along $y$ for $\lambda < 1$ causes the partial decoupling of $q_x$ and $q_y$, but this point needs further studies in the future.

Also in this case, the propagating wave distortion cannot be individuated by one-dimensional case studies or complicated real-world problems, while it is well evidenced by the use of a truly two-dimensional case study in controlled conditions.

[Insert Figure 7 about here]

## 5. Conclusions

Friction decoupling, i.e. the computation of friction vector components making separate use of the corresponding velocity components, is common in staggered grid models of the SWE simplifications (Zero-Inertia and Local Inertia Approximation), due to the programming simplicity and to the corresponding calculations speed-up. Despite the interest in understanding the modifications introduced by friction decoupling, scarce attention has been given in the literature to this topic.

In the present paper, the effect of friction decoupling in flooding models has been studied from the theoretical and numerical point of view. First, it has been found that the decoupling of the friction vector causes the reduction of the computed friction force and rotation of the friction line of action. Second, it has been demonstrated that decoupled-friction models lack of rotational invariance, i.e. the results of the model depend on the alignment of the reference framework. These theoretical results have been confirmed by means of numerical experiments. In particular, it has been found that the friction decoupling causes the distortion of the propagating wave in a radially symmetric experiment, and that the distortion depends on the computational grid alignment. This issue is in agreement with the theory of Cellular Automata where oversimplified rules are used for the exchanges between cells. In addition to these results, it has been found that the celerity of the propagating wave is overestimated in decoupled-friction models. Notably, calibration of the friction coefficient does not solve these issues.

On the basis of the above findings, it is evident that the decoupling of friction vector causes a major loss of credibility of the corresponding mathematical and numerical models, because it is impossible to *a priori* predict how much the decoupled-friction solution will depart from the coupled-friction one. Despite the modest speed-up of decoupled-friction computations, the use of coupled-friction models is preferable in every case.

**Acknowledgments**

The research project "Sistema di supporto decisionale per il progetto di casse di espansione in linea in piccoli bacini costieri" was funded by the Italian Ministry for Environment, Land and Sea Protection through the funding program "Metodologie per la valutazione dell'efficacia sulla laminazione delle piene in piccoli bacini costieri di sistemi di casse d'espansione in linea realizzate con briglie con bocca tarata".

**Appendices**

**A. Rotation of the frame reference for the Shallow water Equations**

In this Appendix it is demonstrated that the Eq. (14) is obtained from Eq. (10) after a rotation of the reference framework. The application of the first and the second of Eq. (13) to the first of Eq. (1) leads to

(A.1) $\dfrac{\partial h}{\partial t}+\cos\theta\dfrac{\partial q_x}{\partial X}-\sin\theta\dfrac{\partial q_x}{\partial Y}+\sin\theta\dfrac{\partial q_y}{\partial X}+\cos\theta\dfrac{\partial q_y}{\partial Y}=0.$

Recalling that $\cos^2\theta+\sin^2\theta=1$, the substitution of the first and the second of Eq. (12) into Eq. (A.1) leads to the first of Eq. (14).

The application of the same procedure to the second of Eq. (1) leads to

(A.2) $\cos\theta\left[\dfrac{\partial Q_X}{\partial t}+\dfrac{\partial}{\partial X}\left(\dfrac{gh^2}{2}+\dfrac{Q_X^2}{h}\right)+\dfrac{\partial}{\partial Y}\dfrac{Q_X Q_Y}{h}+gh\dfrac{\partial z}{\partial X}+g\dfrac{n_M^2 Q_X\|\mathbf{Q}\|}{h^{7/3}}\right]$
$-\sin\theta\left[\dfrac{\partial Q_Y}{\partial t}+\dfrac{\partial}{\partial X}\dfrac{Q_X Q_Y}{h}+\dfrac{\partial}{\partial Y}\left(\dfrac{gh^2}{2}+\dfrac{Q_Y^2}{h}\right)+gh\dfrac{\partial z}{\partial Y}+g\dfrac{n_M^2 Q_Y\|\mathbf{Q}\|}{h^{7/3}}\right]=0.$

Similarly, the application of the rotation procedure to the third of Eq. (1) leads to

(A.3) $\sin\theta\left[\dfrac{\partial Q_X}{\partial t}+\dfrac{\partial}{\partial X}\left(\dfrac{gh^2}{2}+\dfrac{Q_X^2}{h}\right)+\dfrac{\partial}{\partial Y}\dfrac{Q_X Q_Y}{h}+gh\dfrac{\partial z}{\partial X}+g\dfrac{n_M^2 Q_X\|\mathbf{Q}\|}{h^{7/3}}\right]$
$+\cos\theta\left[\dfrac{\partial Q_Y}{\partial t}+\dfrac{\partial}{\partial X}\dfrac{Q_X Q_Y}{h}+\dfrac{\partial}{\partial Y}\left(\dfrac{gh^2}{2}+\dfrac{Q_Y^2}{h}\right)+gh\dfrac{\partial z}{\partial Y}+g\dfrac{n_M^2 Q_Y\|\mathbf{Q}\|}{h^{7/3}}\right]=0.$

If Eqs. (A.2) and (A.3) are summed after the multiplication by $\cos\theta$ and $\sin\theta$, respectively, the second of Eq. (A.14) is obtained. Similarly, if Eqs. (A.2) and (A.3) are summed after the multiplication by -$\sin\theta$ and $\cos\theta$, respectively, the third of Eq. (A.14) is obtained. The proof is complete.

**Figures**

Figure 1. Comparison of coupled and decoupled friction vectors: ratio between the moduli (a); absolute deviation angle (b).

Figure 2. Staggered grid stencil.

Figure 3. Propagation of a circular wave. Free-surface elevation with the Zero-Inertia model at time $t = 60000$ s. Left column, coupled friction: $\Delta s = 3218.6$ m (a); $\Delta s = 1609.3$ m (c); $\Delta s = 804.65$ m (e). Right column, decoupled friction: $\Delta s = 3218.6$ m (b); $\Delta s = 1609.3$ m (d); $\Delta s = 804.65$ m (f).

Figure 4. Propagation of a circular wave. Free-surface elevation with the LInA model ($\lambda = 1$) at time $t = 60000$ s. Left column, coupled friction: $\Delta s = 3218.6$ m (a); $\Delta s = 1609.3$ m (c); $\Delta s = 804.65$ m (e). Right column, decoupled friction: $\Delta s = 3218.6$ m (b); $\Delta s = 1609.3$ m (d); $\Delta s = 804.65$ m (f).

Figure 5. Propagation of a circular wave. Free-surface elevation with the LInA model ($\lambda = 1$) at time $t = 60000$ s. Thin line: coupled friction with $n_M = 0.03$ s m$^{-1/3}$. Thick dashed line: decoupled friction with $n_M = 0.0325$ s m$^{-1/3}$.

Figure 6. Propagation of a circular wave. Free-surface elevation with the LInA model ($\lambda = 1$) and decoupled friction at time t = 60000 s (refined grid): reference framework $Oxy$ (a); reference framework $OXY$ rotated by $\theta = \pi/4$ with respect to the reference $Oxy$. Dashed line: stretching direction of the propagating wave.

Figure 7. Propagation of a circular wave. Free-surface elevation with the coupled-friction LInA model at time t = 60000 s. Left column, $\lambda = 1$: $\Delta s = 3218.6$ m (a); $\Delta s = 1609.3$ m (c); $\Delta s = 804.65$ m (e). Right column, $\lambda = 0.7$: $\Delta s = 3218.6$ m (b); $\Delta s = 1609.3$ m (d); $\Delta s = 804.65$ m (f).

**Tables**

Table 1. Resume of constant time steps and uniform grid sides used in the computations.

Table 2. Run times for the refined grid.

Table 1. Resume of constant time steps and uniform grid sides used in the computations.

| Grid | Panel | Δs (m) | Δt (s) |
|---|---|---|---|
| Coarse | Upper: Fig. 3a,b; Fig. 4a,b; Fig. 7a,b | 3218.6 | 50 |
| Medium | Central: Fig. 3c,d; Fig. 4c,d; Fig. 7c,d | 1609.3 | 12.5 |
| Refined | Lower: Fig. 3e,f; Fig. 4e,f; Fig. 7e,f | 805.65 | 3.125 |

Table 2. Run times for the medium grid

| Model | Run time (ms) | Speedup |
|---|---|---|
| Zero-Inertia with coupled-friction | 18626 | 1.25 |
| Zero-Inertia with decoupled-friction | 14867 | |
| LInA with coupled-friction | 33493 | 1.17 |
| LInA with decoupled-friction | 28673 | |

Figure 1. Comparison of coupled and decoupled friction vectors: ratio between the moduli (a); absolute deviation angle (b).

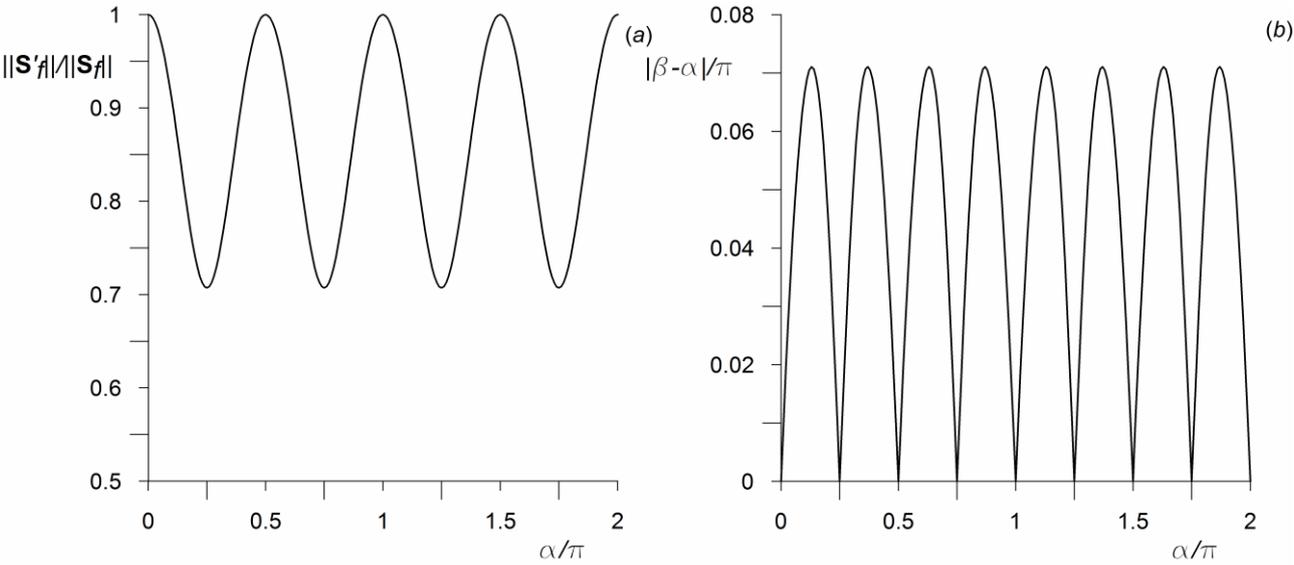

Figure 2. Staggered grid stencil.

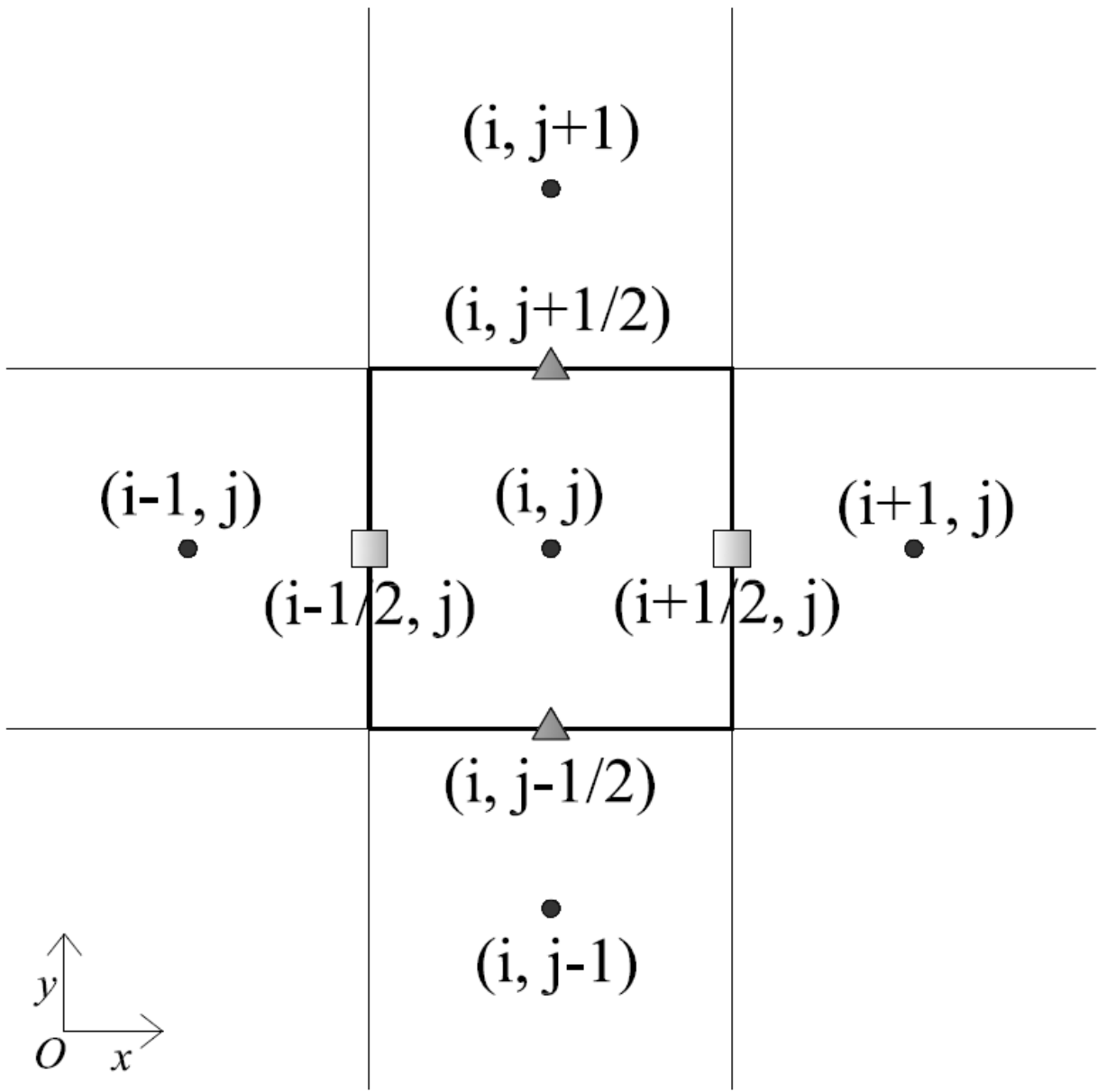

Figure 3. Propagation of a circular wave. Free-surface elevation with the Zero-Inertia model at time $t = 60000$ s. Left column, coupled friction: $\Delta s = 3218.6$ m (a); $\Delta s = 1609.3$ m (c); $\Delta s = 804.65$ m (e). Right column, decoupled friction: $\Delta s = 3218.6$ m (b); $\Delta s = 1609.3$ m (d); $\Delta s = 804.65$ m (f).

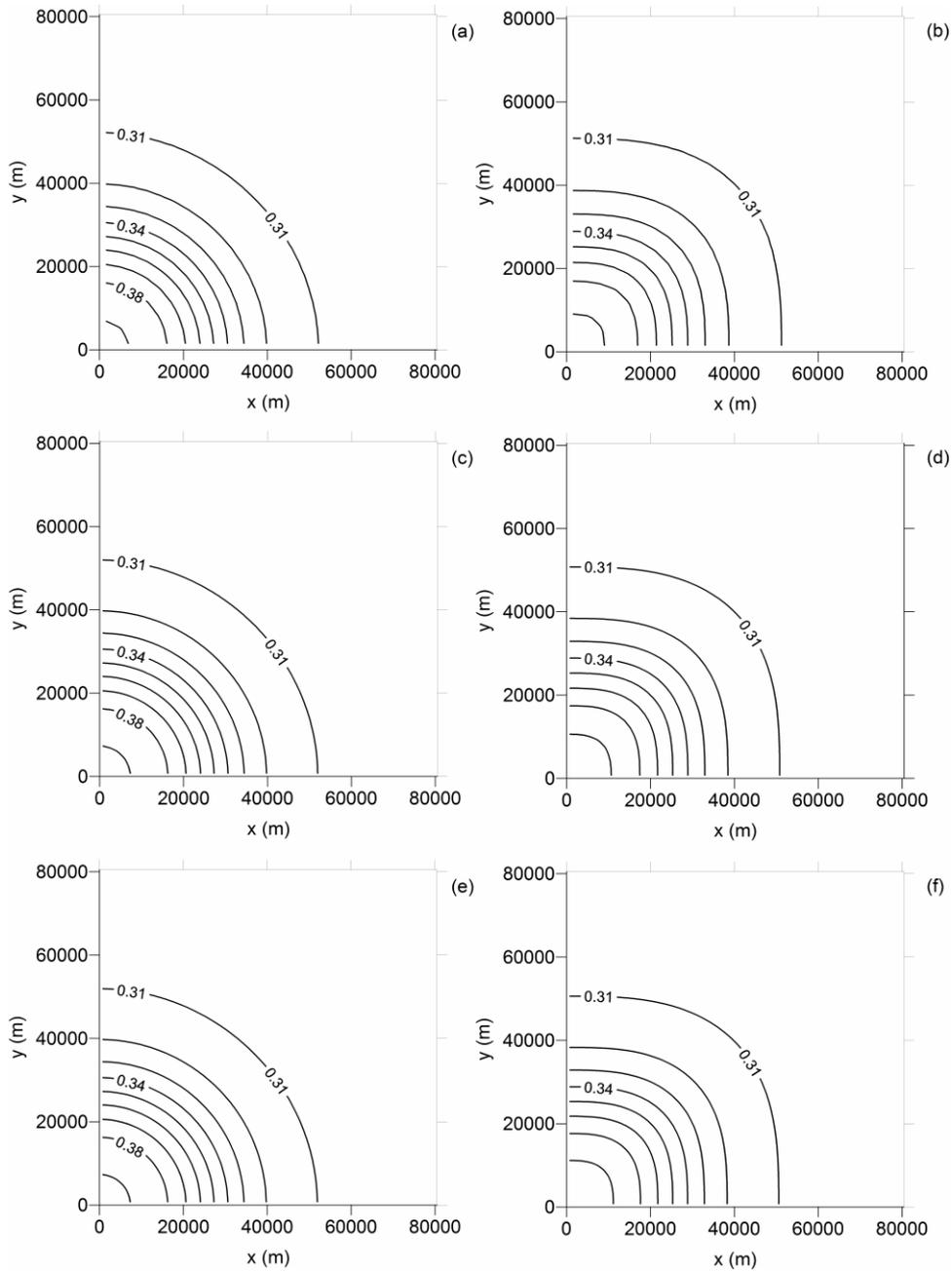

Figure 4. Propagation of a circular wave. Free-surface elevation with the LInA model ($\lambda = 1$) at time $t = 60000$ s. Left column, coupled friction: $\Delta s = 3218.6$ m (a); $\Delta s = 1609.3$ m (c); $\Delta s = 804.65$ m (e). Right column, decoupled friction: $\Delta s = 3218.6$ m (b); $\Delta s = 1609.3$ m (d); $\Delta s = 804.65$ m (f).

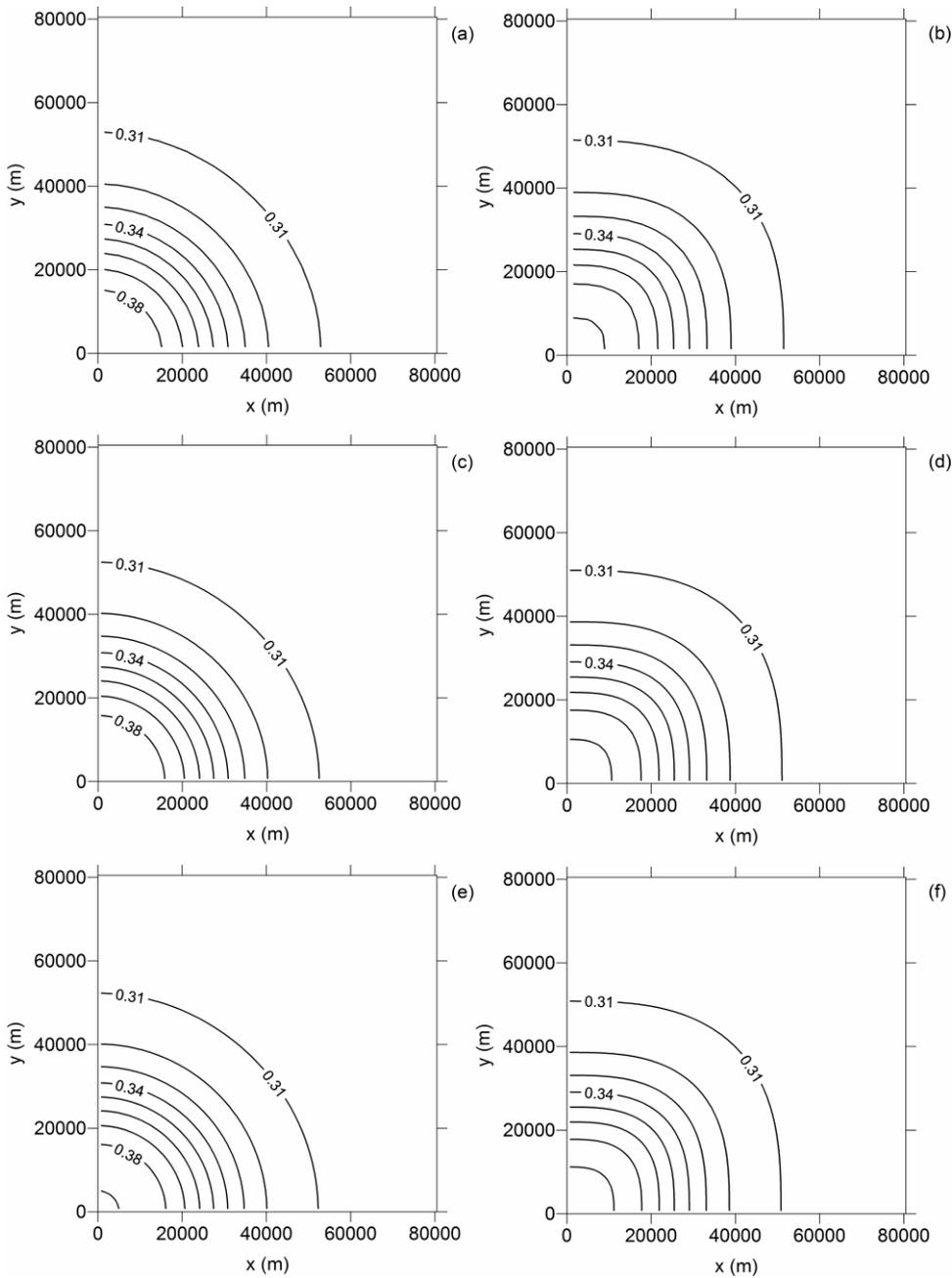

Figure 5. Propagation of a circular wave. Free-surface elevation with the LInA model ($\lambda = 1$) at time $t = 60000$ s. Thin line: coupled friction with $n_M = 0.03$ s m$^{-1/3}$. Thick dashed line: decoupled friction with $n_M = 0.0325$ s m$^{-1/3}$.

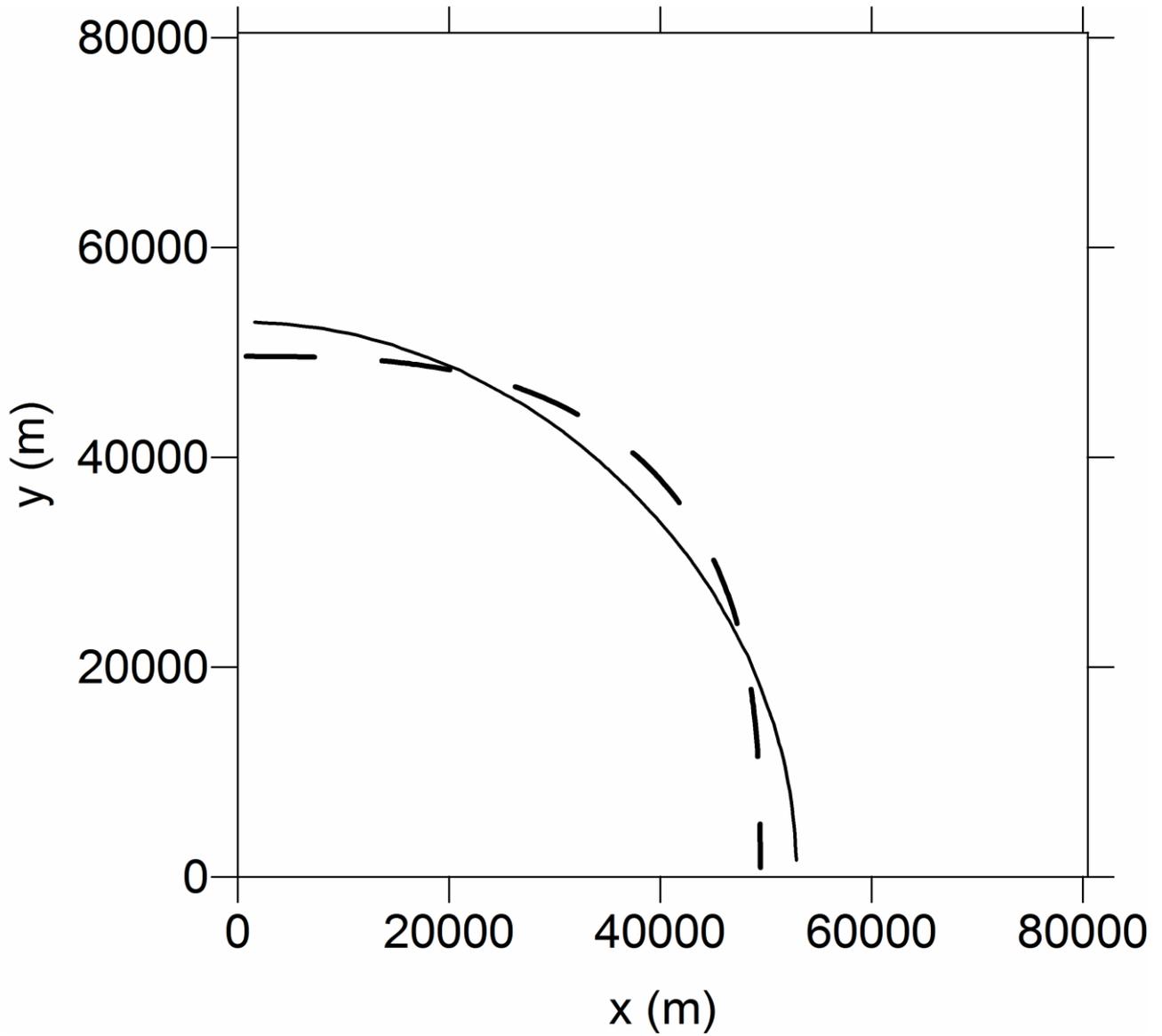

Figure 6. Propagation of a circular wave. Free-surface elevation with the LInA model ($\lambda = 1$) and decoupled friction at time $t = 60000$ s (refined grid): reference framework $Oxy$ (a); reference framework $OXY$ rotated by $\theta = \pi/4$ with respect to the reference $Oxy$. Dashed line: stretching direction of the propagating wave.

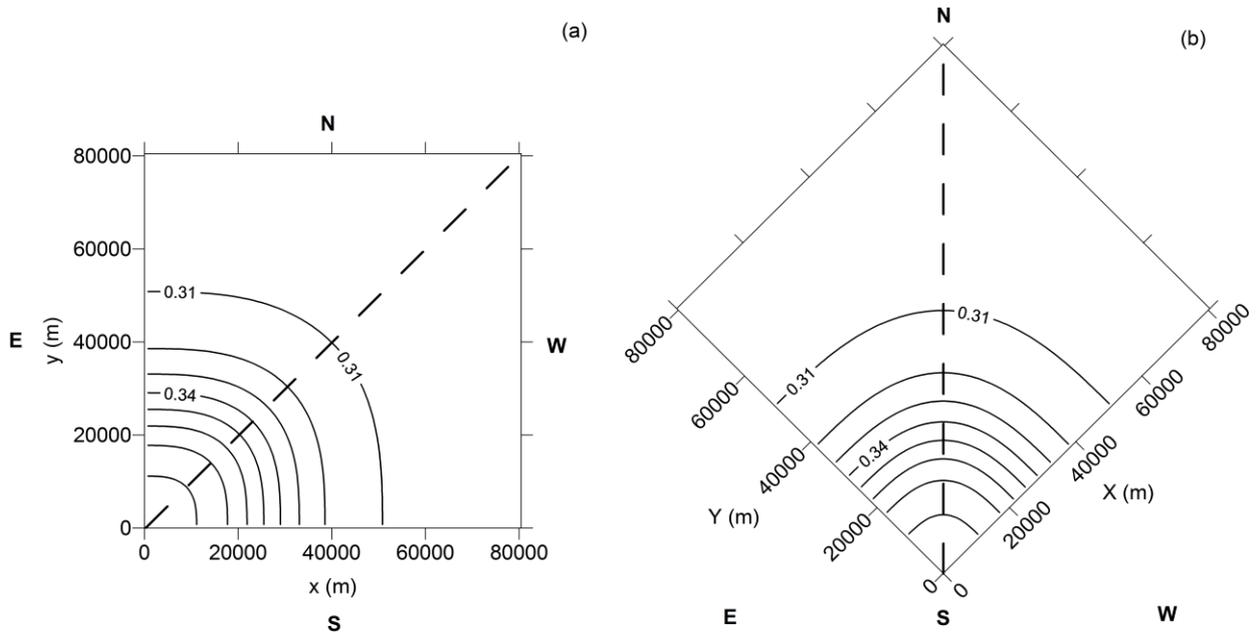

Figure 7. Propagation of a circular wave. Free-surface elevation with the coupled-friction LInA model at time t = 60000 s. Left column, $\lambda = 1$: $\Delta s = 3218.6$ m (a); $\Delta s = 1609.3$ m (c); $\Delta s = 804.65$ m (e). Right column, $\lambda = 0.7$: $\Delta s = 3218.6$ m (b); $\Delta s = 1609.3$ m (d); $\Delta s = 804.65$ m (f).

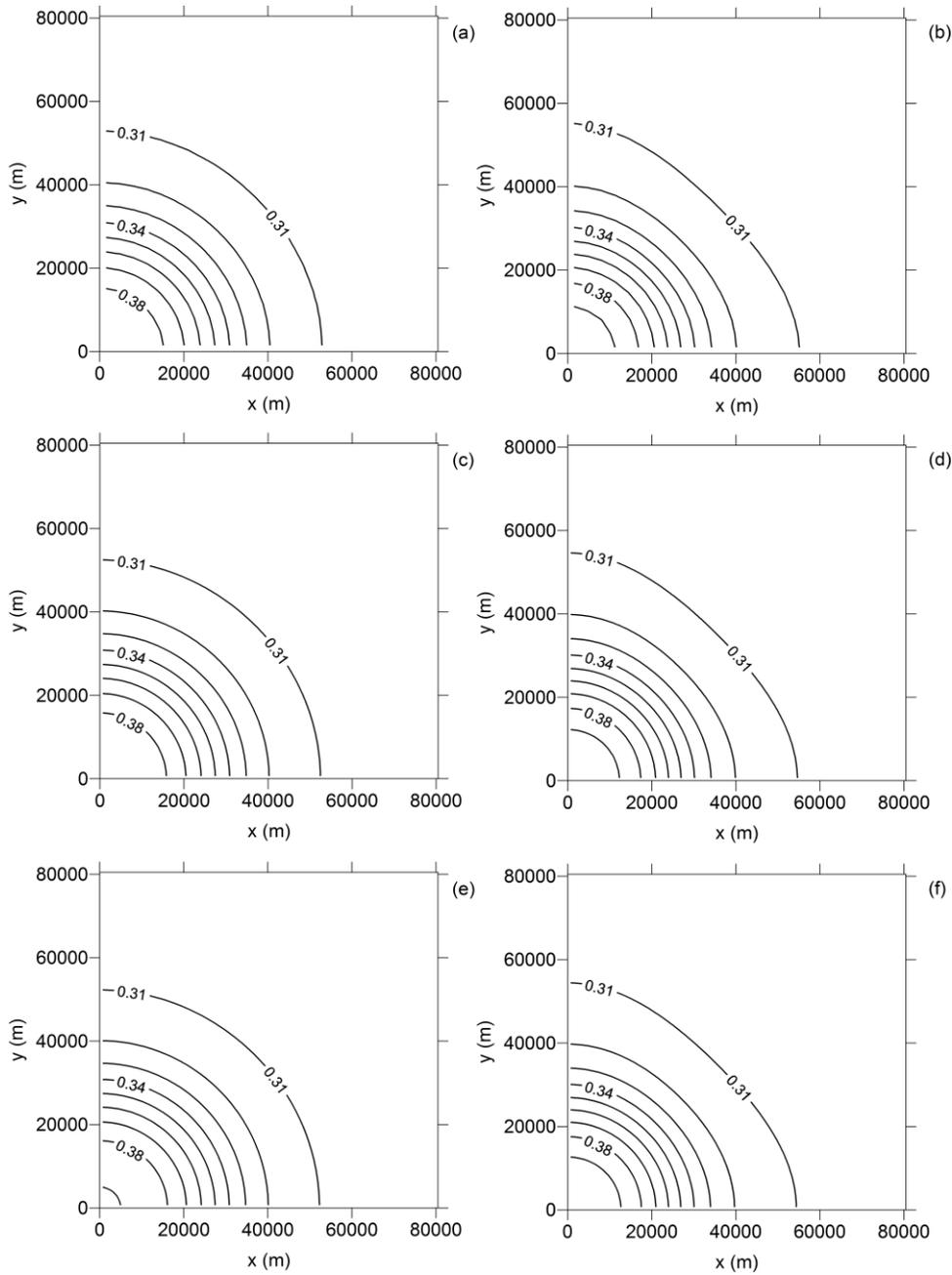